\documentclass[journal]{IEEEtran}
\usepackage{mathtools,cuted}
\usepackage{graphicx}
\usepackage[usenames, dvipsnames]{color}
\usepackage{color,soul}
\usepackage{xcolor}
\usepackage{multirow}
\usepackage{makecell}

\ifCLASSINFOpdf
\else
\fi
\graphicspath{{images_eps/} }

\begin{document}
%
\title{An All-Memristor Deep Spiking Neural Computing {System}: A Step Towards Realizing the Low Power, Stochastic Brain}
%
%
%

\author{Parami~Wijesinghe, Aayush Ankit, Abhronil Sengupta~\IEEEmembership{Member,~IEEE,}
and~Kaushik Roy,~\IEEEmembership{Fellow,~IEEE,}\\ Purdue University, School of Electrical \& Computer Engineering\\ 465, Northwestern Ave, West Lafayette, IN, 47907, USA \vspace{-15pt}
}
\maketitle

\begin{abstract}
Deep `Analog Artificial Neural Networks' (ANNs) perform complex classification problems with remarkably high accuracy. However, they rely on humongous amount of power to perform the calculations, veiling the accuracy benefits. The biological brain on the other hand is significantly more powerful than such networks and consumes orders of magnitude less power, indicating us about some conceptual mismatch. Given that the biological neurons communicate using energy efficient trains of spikes, and the behavior is non-deterministic, incorporating these effects in Deep Artificial Neural Networks may drive us few steps towards a more realistic neuron. In this work, we propose how the inherent stochasticity of nano-scale resistive devices can be harnessed to emulate the functionality of a spiking neuron that can be incorporated in deep stochastic Spiking Neural Networks {(SNN)}. At the algorithmic level, we propose how the training can be modified {to convert an ANN to an SNN} while supporting the stochastic activation function offered by these devices. We devise circuit architectures to incorporate stochastic memristive neurons along with memristive crossbars which perform the functionality of the synaptic weights. We tested the proposed All Memristor {deep stochastic SNN} for image classification and observed only about $1\%$ degradation in accuracy with the ANN baseline after incorporating the circuit and device related non-idealities. We witnessed that the network is robust to certain variations and consumes $\sim 6.4\times$ less energy than its CMOS counterpart.
\end{abstract}


%
\IEEEpeerreviewmaketitle

\section{Introduction}\label{intro}
%
%
%
%
\IEEEPARstart{E}{ven} though the exact mechanisms of communication between biological neurons still remain unknown, it has been shown experimentally that neurons use spikes for communication and the nature of the firing of neurons (spike generation) is non-deterministic \cite{2,3,4}. By conserving energy via spike based operation \cite{4post}, the brain has evolved to achieve its prodigious signal-processing capabilities using orders of magnitude less power than the state-of-the-art supercomputers. Therefore, with the intention to pave pathways to low power neuromorphic computing, much consideration is given to more realistic artificial brain modeling \cite{5}. The inception of the Spiking Neural Networks (SNN) concept is a consequence of above. It has recently emerged as an active area of research owing to its resemblance of the ``actual human brain'' \cite{1}.

In a spiking neural network, the communication between neurons take place by means of spikes. The information is typically encoded in the rate of occurrence of spikes. Different learning schemes have been proposed over the past, and Spike Time Dependent Plasticity (STDP) based learning is widely used due to the consistency of the concept with experimental statistics \cite{6}. However, the STDP learning is typically limited to a network with a single layer of excitatory neurons and a single layer of inhibitory neurons \cite{7}. The aptitudes of such a single fully connected layer spiking neural network is limited when compared with the high recognition performances offered by deep ANNs. Up to date, deep ANNs have given the best performance with respect to classification accuracy. For an example, SENet which won the 2017 ILSVRC, is a deep Convolutional Neural Network (CNN) with the reported lowest top 5 error (the correct class is not within the top 5 selection of classes according to the network output) of 2.251\% on ImageNet data set \cite{8}. However, such networks require huge power and time if a von-Neumann computer is to be used for computation. For an instance, SE-ResNet requires power for $\sim$3.2GFLOPS (number of operations per second) \cite{8}. 

As an effort of embedding the classification accuracy of such ANNs with the spike based low power operation of SNNs, numerous research efforts have been focused on crafting Deep Spiking ANNs \cite{dsnn}. One of the interesting mechanisms of executing the above is by exploiting certain optimization techniques to convert a fully trained deep ANN to an SNN \cite{9,10}. The work suggested in \cite{9} outperforms all previous SNN architectures to date on MNIST database. Despite the existence of Deep SNNs in the algorithmic level, minimal consideration is dedicated towards devices and realizing such algorithms in hardware level. 

With the intention to reduce the energy consumption of the powerful Deep ANNs while preserving the biological plausibility, we propose a non-deterministic, memristive device based hardware architecture for a {Deep Stochastic SNN}. Memristors have been widely used in literature as synapses in neural networks \cite{12,13}. The multi-level storage capability has made the memristor an ideal candidate for the synapses in a neural network. Even though this multi-level behavior of the memristors seems appealing to emulate the behavior of an analog neuron as well (different voltage write pulses (inputs) result in different memristor resistances (output); there is an upper and a lower bound for the resistances; this signals similar functionality of a thresholding function), reliability concerns might arise due to the inherent stochasticity. This stochasticity in memristors has been experimentally shown \cite{14,15} and the statistical measures suggest that the switching probability of these devices can be predicted. For example, the switching times follow a Poisson distribution for Silver/amorphous Silicon/poly Silicon ($Ag$/$a-Si$/$p-Si$) based devices.

Memristors offer a variety of favorable features such as higher write-erase cycles ($10^{12}$ \cite{16}), higher yield, CMOS compatibility, lower area $etc$. Despite these benefits, high programming voltages and long pulse durations, \cite{17} or other feedback write mechanisms \cite{17post} are mandatory to ensure the switching of the devices, for applications such as memory, that require very low failure rates. Rather than trying to reduce such non-deterministic effects, in this work we propose an effort to embrace the stochasticity in an efficient way, with the ambition to go towards a more realistic neuron. We propose the memristor as a probabilistic switch to represent the stochastic neuron in a supervised {deep stochastic spiking neural network}, and memristive crossbar arrays with multi-bit capability to represent the `inner product' computation between the incoming spikes and the synaptic weights. We introduce this structure as `All-Memristor' neural network due to the fact that the two main functionalities of a neural network are being taken care of by memristors. We elaborate how the ANNs can be trained, in order to incorporate a stochastic memristor as a neuron. The gradient descent based backward propagation scheme must be modified to account for the probabilistic function which may be different from standard activation functions (ReLU, sigmoid $\big(\dfrac{1}{1+e^{-x}}\big)$, $etc.$) of a neuron. We will further elaborate certain favorable features accompanied by memristors that makes it suitable to emulate a stochastic neuron. We propose circuit architectures that can be used to realize the proposed All-Memristor network. Then the impact of certain variations towards the accuracy of the network is explored. Finally we compare the energy consumption of the All-Memristor based network with the CMOS counterpart.

Even though the possibility of harnessing the inherent stochasticity of the memristor for neuromorphic computations has been mentioned previously \cite{18, hu, IBM}, the complete analysis of it for deep stochastic SNNs has not yet been studied. Further, stochastic integrate and fire neurons (with a focus on devices) have been proposed in literature \cite{20,21} for unsupervised learning SNNs and they are different from this work, where we have specifically designed the neuron to suit deep supervised neural networks which are capable of performing complex tasks with better accuracy \cite{9}. We {provide complete analysis of hardware neural network} with memristors, to which four key features of the brain; high accuracy, low power, spike based information transferring, and stochasticity are embedded. Note that this work is based on \cite{aSNN}, which explored the design of such deep stochastic SNNs for current controlled magnetic devices. This work extends the concept for memristive devices which are voltage controlled devices and therefore requires a re-thinking of the design.

\begin{figure}[b!]
\centering
\vspace{-15pt}
\setlength{\abovecaptionskip}{-5pt}
\includegraphics[width=3.5in]{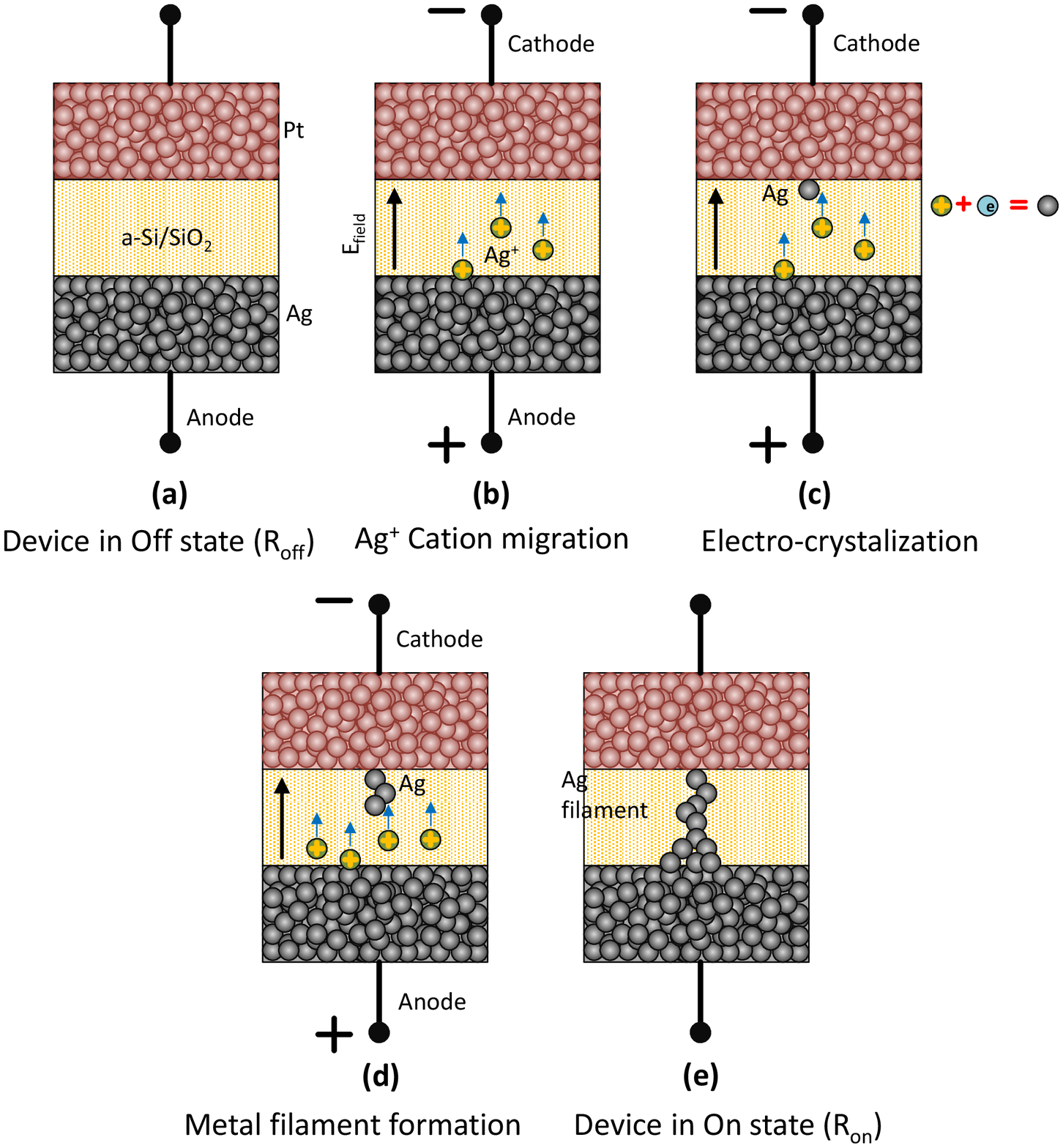}
\caption{The standard SET operation for an ECM type memristor (a)`Off' state of a memristor has higher resistance due to the sandwiched insulating material. (b) To `SET' the device, positive voltage must be applied to the active electrode with respect to the inert electrode. Ag cations start traveling towards the inert electrode due to the $E_{field}$. (c),(d) The $Ag^+$ ions get combined with electrons and crystalize forming a metal filament. (e) Once a full metal filament is formed between the two electrodes, the resistance of the device lowers.}
\label{Device}
\end{figure}

\section{Memristors as synapses and neurons}

Nanoscale resistive devices have been extensively studied as a leading candidate for non-volatile memory \cite{mem}, reconfigurable logic \cite{logic}, and analog circuits \cite{analog}. The possibility of using different types of memristors in different types of neural networks has also been explored. The ability to change resistance due to voltage pulses makes the memristor a better candidate for STDP learning \cite{12}. Some memristors show unstable intermediate resistance states which are suitable as synapses to represent the short-term memory and long-term memory functionality \cite{13}. 

In this work, our network has been trained as an ANN to obtain the proper synaptic weights. These synaptic weights can be represented by conductances (refer to section {\ref{cctimp}} for more details on this). The memristors that act as synapses in ANNs should ideally be able to have ‘any’ conductance value. Therefore, we are using the multi-bit capability of the memristors to represent the synaptic functionality of the neural network. As mentioned previously, in this work, we are also representing the functionality of neurons using memristors. For this, we have selected the stochastic binary switching behavior of the memristors.

The typical nano scale resistive device is based on a metal-insulator-metal (MIM) structure. The resistance change in these devices can be attributed to the formation of a conductive filament inside the insulator ($Ag$, amorphous $Si$ ($a-Si$), $Pt$ based devices), change in the phase due to Joule heating and cooling (chalcogenide based devices), or field assisted drift/diffusion of ions ($TiO_2$ based devices). These processes have shown to be random in nature. For this work, we are considering the $a-Si$ based metal filament formation devices (Electrochemical Metallization devices or ECM devices\cite{23,24}) due to multiple reasons as explained below. However, it must be noted that we have selected this particular type of device as an example for a memristor to show the applicability of it for the deep stochastic SNNs. A different type of a memristor can suit better in different contexts (example: network accuracy, power consumption $etc.$). For example, the HfO$_x$ based devices have higher endurance and lower switching time \cite{hfo}. $Ag/AgSiO_2$ devices reset after a certain time period eliminating the requirement for resetting \cite{26} (refer to section {\ref{result}} for more details). 

$a-Si$ memristors typically have very high resistance ($\sigma \sim 3\times10^{-5}\Omega cm^{-1}$ at $310$K). When power consumption is considered, it is better to have high resistances in memristive crossbars. It is also possible to adjust the lower resistance of the device to suit the constraints of the crossbar driving sources. This can be done by tuning the $a-Si$ growth conditions ($R_{ON}$ can be varied from $\sim100$M$\Omega$ to $10$k$\Omega$) during the PECVD or LPCVD deposition processes \cite{spec}. The ON-OFF ratio of the device is high ($\sim 10^7$) as well \cite{14}. Having a higher ratio implies the higher reliability of programming a single state under variations in multi-bit configurations {\cite{multi}}. This is also better in terms of sensing if a memristor has switched or not, in the context of a neuron. Unlike some types of memristors, the $a-Si$ memristors require high write voltages. For an example, the nanoporous $SiO_x$ based devices have very low forming voltages ($\sim1.4$V) \cite{rice_uni_multibit}. Having smaller operating voltages in memristive crossbars require sophisticated sensing mechanisms. This also signals a reliability concern for the nanoporous multi-bit $SiO_x$ devices, while operating in a crossbar with larger driving voltages. In contrast, the $a-Si$ based memristor can operate at $2$V reading voltages without disturbing the device resistance \cite{spec}.

The possibility of the $a-Si$ ($Ag$ based) devices to act both as a muti-bit storage and as a stochastic switch {\cite{14}} is beneficial for this work since it can act both as a neuron and a synapse. As it will be explained in the next section, in an $a-Si$ based memristor, a metal filament is formed between the two contacts when going from OFF to ON state. The $Ag$ particles goes to the defect sites inside $a-Si$ and creates this filament. Depending on the number of defects, the I-V characteristics of the device shows multiple abrupt jumps in currents \cite{14}. Devices with lower lengths will have lower number of defects and are much suitable for binary switching applications (which is the neuron in our work). When programming for multi levels, current/voltage must be controlled properly. It has been experimentally shown the possibility to store 8 levels of resistances (3-bit storage) {\cite{14}} using the same write voltage pulse and different series resistors ($R_s$) to control the current. Each $R_s$ resulted in different final resistance values of the memristor. Given the fact that a $30$nm device can store 3-bit levels, and the resistance of memristors are proportional to the device length, it can be fairly assumed that a 60nm device can store 4-bit levels. It has been shown that nanoporous $SiO_x$ memristors can store up to 9-bits \cite{rice_uni_multibit} per cell. Such devices can give higher accuracy if used in hardware ANNs to represent the synapses. 

Even though the stochasticity is helpful to represent the functionality of a stochastic neuron, it may not be beneficial for the supervised learning scheme being explored in this work for multi-bit synapses. Due to stochasticity, programming using a single voltage pulse with the selected control resistor may not guarantee the device transferring to the expected resistance level. Furthermore, due to variations, the value of resistance at each level may change. In a subsequent section, we show that the accuracy degrades significantly when the conductance variations are larger than $25\%$. Therefore, the need for better memristor programming schemes arise. It has been experimentally shown that $TiO_2$ based filament type memristors can be programmed to have a required resistance (within $1\%$ accuracy) using a novel programming scheme, despite the variability {\cite{write1}}.

In the next section, the aforementioned stochasticity in memristors that we have incorporated in neurons will be explained in the device level.

\begin{figure}[b!]
\centering
\vspace{-15pt}
\setlength{\abovecaptionskip}{-2pt}
\includegraphics[width=3.5in]{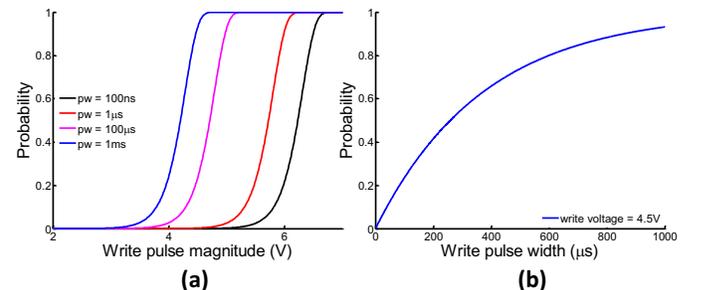}
\caption{(a) The switching probability of a memristor with varying magnitude of the pulse. (b) The Switching probability of a memristor with varying pulse width}
\label{PvsV}
\end{figure} 

\section{Stochasticity in memristor devices}\label{neuron}

Despite the copious favorable features offered by memristive devices, the stochasticity of changing its state has induced reliability concerns. To make the memristor a deterministic device in order to appropriate it for applications such as non-volatile memories, reconfigurable logic $etc.$, significant consideration must be provided to the operating region of the devices. As an example, the $TiO_2$ based memristive devices have a typical threshold voltage of 1V in order to `SET' the device \cite{22} for memory applications. This is the magnitude of the voltage write pulse that provides a higher confidence (ex: $>0.99$) of writing a logic 1. If a writing pulse of 0.5V is applied instead of 1V, the device may change its state with a certain probability which is less than 1. It is evident that increasing the reliability comes at the cost of high power consumption. Our work is an effort to utilize the stochastic behavior of the nanoscale resistive devices while operating in the low-power non-deterministic regime.

The ECM devices consist of an insulating membrane ($a-Si$, $SiO_2$, $Al_2O_3$) sandwiched between two active($Ag$) and inert electrodes($ Pt, Ti$). When the device is in its higher resistance ($R_{off}$) state, it is considered as storing a logic `0' (in memory). When writing a `1' to the device (SET or `turning on'), a positive voltage must be applied to the active electrode with respect to the inert electrode. At this point, the active electrode dissolution transpires and cations from the active electrode start migrating towards the inert electrode where it gets electro-crystalized, forming a metal filament (termed as the `forming process'\cite{24}). Once a full metal filament has grown between the two electrodes, there is a sudden drop in resistance. The aforementioned process is graphically explained in Fig. \ref{Device}. The low resistance ($R_{on}$) stage of the device is assigned to logic `1'.

\begin{figure}[t]
\centering
\setlength{\belowcaptionskip}{-25pt}
\includegraphics[width=3.5in]{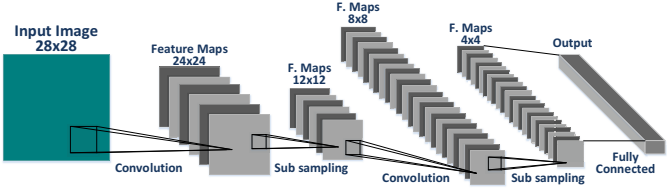}
\caption{The typical structure of a convolutional neural network. There are two main sections in a CNN in terms of functionality. The convolutional layers (followed by subsampling to reduce the large number of computations) perform the feature extractions from an input (ex: image), and the fully connected layer classifies the inputs depending upon the extracted features.}
\vspace{-10pt}
\label{cnn}
\end{figure}

The formation of the filament is a highly voltage bias dependent process. The anode metal particle hopping rate is given by \cite{24}
\begin{equation}
\Gamma = \dfrac{1}{\tau}=\nu e^{-E_a^{\prime}(V)/k_BT}
\label{hopping_rate}
\end{equation}

where $k_B$ is the Boltzmann’s constant, $T$ is the temperature, $\tau$ is the characteristic switching time, and $\nu$ is the attempt frequency. $-E_a^{\prime} (V)$ is the bias dependent activation energy. The time required for the formation of the metal filament is shown to follow a Poisson distribution \cite{14}. The probability of switching within the next $\Delta t$ duration after a $t$ amount of time can be defined as
\begin{equation}
P(t) = \dfrac{\Delta t}{\tau}e^{-t/\tau}
\label{pdf}
\end{equation}

The dependency between the characteristic switching time and the voltage of the write pulse is given by 

\begin{equation}
\tau(V) = \tau_0e^{-V/V_0}
\label{tau}
\end{equation}

\vspace{-10pt}
\begin{equation}
{\tau_0 = \dfrac{1}{\nu}e^{E_a/k_BT},\quad V_0 = 2nk_BT/q}
\label{tau0}
\end{equation}

where $E_a$ is the activation energy at zero voltage bias, $n$ is the number of anode metal particle sites, $q$ is the charge of an electron.

If a particular write voltage pulse is applied on the memristor, according to above equations, it can be noted that the switching probability depends on two key factors. 

\begin{enumerate}
\item The magnitude of the pulse.
\item The width of the pulse
\end{enumerate}

Fig. \ref{PvsV} (a) shows how the magnitude of the write pulse affects the switching probability curve and Fig. \ref{PvsV} (b) shows the effect of the width of the write pulse. For a rate based spiking neural network, if a memristor must be incorporated as a spiking neuron, the width of the spikes must ideally be the same (variations might be present and their effect is analyzed in the results section). Therefore, the magnitude of the pulses must be controlled to bring the memristor to its stochastic operating regime. From equation (\ref{pdf}), the cumulative probability of switching when a voltage $V$ is applied on the memristor for a $t$ amount of time is
\begin{equation}
{P = 1-e^{-t/\tau} = 1-exp(-\dfrac{t\nu}{e^{E_a/k_BT}}e^{Vq/2nk_BT})}
\label{cdf}
\end{equation}

Once the write time $t$ is selected for the network, all the parameters in the above function is fixed (i.e., $P = f(V)$).

\section{{Deep Stochastic Spiking Neural Networks}}\label{deepnn}

\subsection{Convolutional neural networks basics}

CNNs have multiple hidden layers of neurons between the input and output layers. For an example, the CNN in Fig. \ref{cnn} has 2 convolution layers, two subsampling layers and one fully connected layer. Each convolution and fully connected layer involves calculating the summation of some weighted inputs and then sending the outcome of it through an activation function. This output is fed as an input to the next layer. Calculating the set of synaptic weight values is called training and stochastic gradient descent method is usually used to back-propagate the error at the output and update the weight values. Typical activation functions for a CNN include sigmoid function, tan hyperbolic function and rectified linear function. The activation function for the stochastic neuron in this work is a probabilistic function as will be described in the next section.

\subsection{Stochastic neurons}

Let us first consider an analog neuron with an activation function $f$. The input to the neuron is the weighted summation of the set of outputs from the previous layer. The output of the neuron can be given as
\begin{equation}
{y = f(x\cdot w)}
\label{ann_neuron_out}
\end{equation}
where $x$ is the output vector from the previous layer and $w$ is the set of synaptic weights. The output varies between $0$ and $1$. Therefore, $x$ can be in the form of $x =[0,1]^N$ with $N$ being the number of fan-in neurons. In contrast, the neurons in spiking neural networks, communicate in terms of Poisson spike trains. $i.e.$, instead of the analog input vector $x$, we would have $\tilde{x}(t) = \{0,1\}^N$ where $1$ represents a spiking event and $0$ represents a non spiking event. In an integrate and fire neuron or leaky integrate and fire neuron, the activities at the inputs are integrated over time until the accumulated value (membrane potential) reaches a certain threshold value. Once this threshold value is crossed, the neuron will produce an output spike (neuron fires) and reset the membrane potential. The stochastic spiking neuron that is being discussed in this work does not temporally accumulate the spiking activities until it reaches a predefined threshold. Instead, it incorporates a probability function that observes the spiking activities at the input from the pre-layer neurons during a time step, and produce a spike with a certain probability that depends on the weighted summation of these activities. Throughout this document, the `spiking neuron' term refers to the above context.

\subsection{ANN to stochastic SNN conversion}

In this section, {we elaborate on the conversion of an ANN to an SNN and its associated error}. We observed a similar explanation in \cite{aSNN} for a standard sigmoidal activation function. We decided to include the evaluation here since the activation function of our interest is not a standard sigmoid. 

{There are two types of spiking networks in terms of the way the data is encoded in the pulses. One method is encoding the information in the exact time a spike occurs. In this work, we are considering the second method where the information is included in the rate of the spikes. When converting an ANN to such an SNN, the analog input of an ANN must be rate encoded as a Poisson spike train. The expected value of the input spike events can be elaborated as (for $N = 1$)}

\begin{equation}
\big<\tilde{x}(t)\big> = \dfrac{\sum_{t}{\tilde{x}(t)}}{T} = x
\label{expected_ip}
\end{equation}

where $T$ is a sufficiently large number of time steps. Let us assume that we have considered a probabilistic activation function similar to the analog activation function $f$ (or in other words, consider that the ANN was trained with a function similar to the probability curve $f$ of a device). Here the neuron gives an output spike with a certain probability defined by $f$ depending upon the input events (spiking/not spiking). When there is a spike at time $t$, $\tilde{x}(t)=1$. The corresponding probability of getting a spike at the output is {$\tilde{y}(t)=f(\tilde{x}(t)\cdot w)=f(w)$}, where $w$ is the synaptic weight. Similarly, when there is no spike at time $t$, $\tilde{x}(t)=0$ and the probability of getting a spike at the output is $\tilde{y}(t)=f(0)$. The expected output can be elaborated as 
\begin{equation}
{\big<\tilde{y}(t)\big> = x\cdot f(1\cdot w)+(1-x)\cdot f(0\cdot w)}
\label{expected_op}
\end{equation}

As we explained in section {\ref{neuron}}, the probability of switching a memristor with a voltage pulse of constant pulse width and the input {$(x\cdot w)$} encoded as the magnitude, takes the form of {(for the exact relationship, refer to equation ({\ref{cdf}}). We have refrained from using the extra constants for simplicity of elaboration and understanding. These constants does not affect the concepts that are being discussed in this section) }
\begin{equation}
{f(x\cdot w)= 1- exp(-e^{x\cdot w})}
\label{neuron_prob}
\end{equation}

Therefore the expected value of the output is 
\begin{equation}
{\big<\tilde{y}(t)\big> = x\cdot \Big(1-exp\big(-e^w\big)\Big) +\big(1-x\big)\big(1-e^{-1}\big)}
\label{expected_op_simple}
\end{equation}

\begin{figure}[b!]
\vspace{-5pt}
\centering
\setlength{\abovecaptionskip}{-2pt}
\includegraphics[width=3.6in]{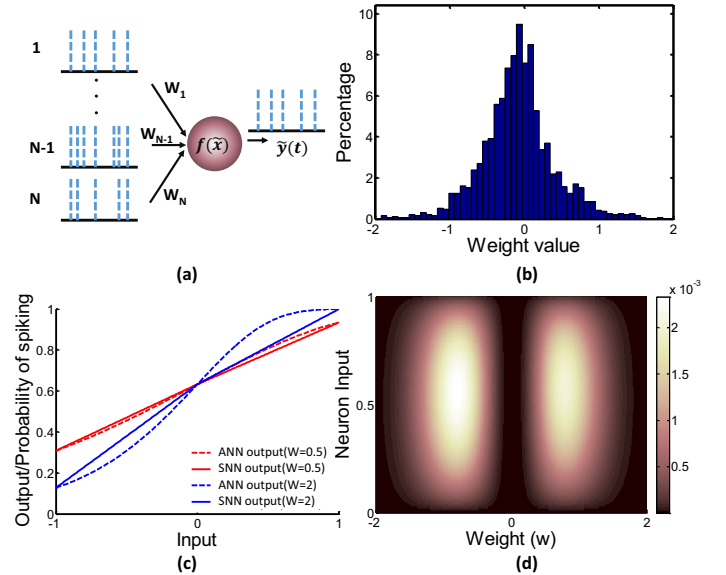}
\caption{(a)The stochastic spiking neuron structure. Incoming inputs are spikes. Depending upon the weighted summation, of spikes, the occurrence of a spike at the output will be determined according to the probability function $f$. (b) The distribution of weights of a network when trained according to a device level probabilistic curve assuming an analog behavior. (c) The expected value of the SNN neuron output for different expected values at the input, and ANN neuron output for different inputs. We have considered two weight values; 0.5 and 2. Negative inputs refer to the cases where the synaptic weight value is negative (d) The ANN to SNN conversion error distribution. Maximum error appears close to 0.5 input spike rate and $|w|=1$. However, the error value is not very significant.}
\label{error}
\end{figure}

For an ideal ANN to SNN mapping (one to one mapping), this expected value must be similar to {$f(x\cdot w)$}. However, as the above equation suggests, $\big<\tilde{y}(t)\big>$ takes a linear form with input $x$. As explained in \cite{aSNN}, the difference between {$f(x\cdot w)$} and $\big<\tilde{y}(t)\big>$ grows with the increasing weight value (Fig. \ref{error} (c)). However, according to the distribution of weights illustrated in Fig. \ref{error} (b) (for a deep ANN trained with the activation function $f$), all the weight values come under the window of $|w|<2$. We can now get an estimate for the error when mapping the ANN to an SNN, assuming the probability of having any spiking rate $\big<\tilde{x}(t)\big>=x=[0,1]$ is equally likely (uniform distribution). Fig. \ref{error} (d) shows the error when we consider having a weight value in the range $|w|<2$ according to the distribution in Fig. \ref{error} (b).

\subsection{Training the ANN before converting to an SNN}

As mentioned in the previous section, we use an activation function similar to the switching probability curve of a memristor given by equation (\ref{neuron_prob}). The weight update rule should change according to this activation function. The stochastic gradient descent weight update rule is as follows

\begin{equation}
\Delta w_{ij}=-\eta \dfrac{\partial{E}}{\partial w_{ij}}=-\eta \dfrac{\partial{E}}{\partial o_j}\dfrac{\partial{o_j}}{\partial net_j}\dfrac{\partial{net_j}}{\partial w_{ij}}
\label{SGD}
\end{equation}

\begin{equation}
o_j =f(net_j)
\label{neuron_out}
\end{equation}

Where $E$ is the cost function that must be minimized for a given input (preferably the squared error at the output). $o_j$ is the output of the $j^{th}$ neuron, $net_j$ is the weighted summation of inputs coming into the $j^{th}$ neuron and $\eta$ is the learning rate. The term $\dfrac{\partial{o_j}}{\partial net_j}$ changes according to the following equation due to the choice of our device defined activation function.
\begin{equation}
\dfrac{\partial{o_j}}{\partial net_j} = (o_j-1) ln(1-o_j)
\label{dout_dnet}
\end{equation}

The bias values in the network are considered to be constant and do not get updated during training. The constant value corresponds to the probability of switching at the output of the neuron during an event of `no spike'. Any output probability during a no spike event can be selected by properly adjusting this bias value.

\section{`All memristor' stochastic SNN architecture}\label{cctimp}

\begin{figure}[t]
\centering
\includegraphics[width=3.5in]{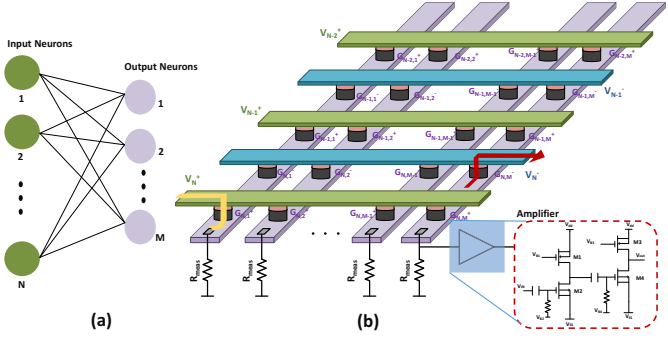}
\caption{(a) A schematic of a fully connected layer in a neural network with N input neurons and M output neurons. (b) The crossbar structure that represents the inner product functionality between the incoming spikes and the synaptic weights}
\vspace{-10pt}
\label{xbar}
\end{figure}

In this work, we consider a deep spiking convolutional neural network which is trained offline, using the switching probability curve explained in the previous section. All the synaptic weights are realized by the conductance of multi-level memristors (4-b discretization levels were used for this case \cite{25, cml, aSNN}. A multilevel writing scheme was proposed in \cite{multi} for TiO$_2$ based memristors using the model in \cite{bio}. The authors claim that the system can write any number of levels given that the on/off ratio is high). The spike trains are short voltage pulses. The inner product between the incoming spikes and the synaptic weights at time $t$ can be efficiently calculated by using a crossbar structure. Let $V(t) = \{0,1\}^N$ be the incoming voltage spikes from $N$ neurons towards the $N\times M$ crossbar ($N$ pre-layer neurons, $M$ post-layer neurons in a fully connected structure). If a conductance value in the crossbar is $G_{i,j}$, then the inner-product between the voltage pulses, and the conductances of the memristors connected to the $j^{th}$ metal column, is the current that flows through the $j^{th}$ metal column itself ($I_j(t)$).
\begin{equation}
{I_j(t)=V(t)\cdot[G_{1,j},G_{2,j}...G_{N,j}]^T}
\label{I_vs_V}
\end{equation}

Ideally, the above value must be converted to a proportional voltage that can bring the memristor to the `stochastic regime' as explained previously. This can be done by sending the above current through a resistor and appropriately amplifying the voltage across it. However, incorporating such measuring resistors ($R_{meas}$) cause non-ideal inner products \cite{aSNN}. Therefore, the measuring resistance must be made considerably small with respect to the values of other resistors that emulate the synaptic functionality. A crossbar coupled with measuring resistors is shown in Fig. \ref{xbar}. Simple low power amplifiers can be incorporated to amplify the voltage across the measuring resistor (Fig. \ref{xbar}) as required. The input impedance of the amplifier is very large. The output impedance is comparatively smaller than the off-resistances of a memristor. The output voltage of the amplifier is biased to give the same probability that the network is trained for (refer to the explanation in section {\ref{deepnn}} D) during an event of no spike. The negative weights are realized by conditionally selecting between positive and negative voltages as shown in Fig. \ref{xbar}. For example, if the weight is negative at the $(i, j)$ cross-point in the cross bar, then the memristor between the $i^{th}$ positive metal row and $j^{th}$ metal column is turned off and vice-versa .

\begin{figure}[b]
\centering
\vspace{-5pt}
\includegraphics[width=3.5in]{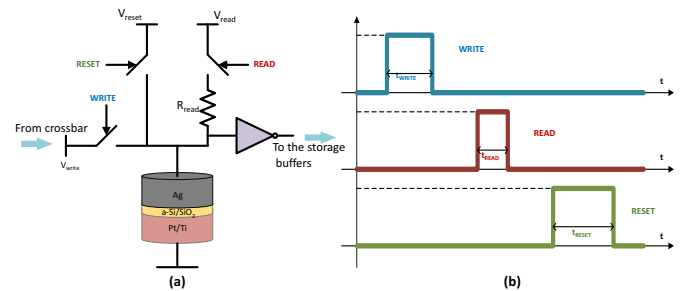}
\caption{(a)The stochastic memristor neuron (b) The temporal variation of write, read, and reset control signals within a single time step }
\label{readwritereset}
\end{figure}

\begin{figure}[t]
\centering
\includegraphics[width=3.5in]{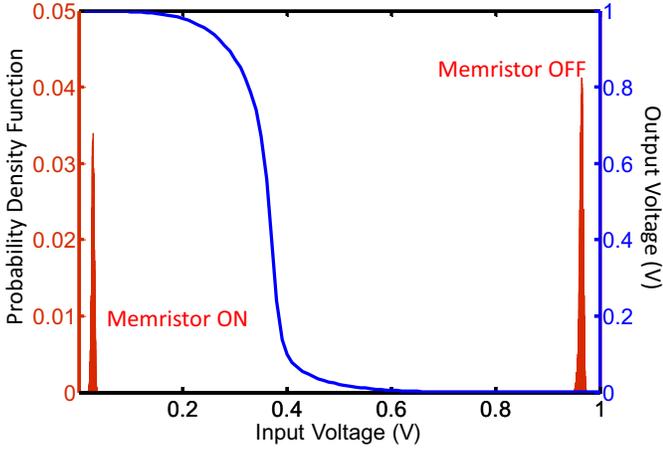}
\caption{{The input and output voltage characteristics of the inverter in the read circuit. The input voltage changes due to variations in ON and OFF resistances (with $\sigma = 20\%$) are shown in red.}}
\label{inv}
\vspace{-10pt}
\end{figure}

Each time step of operation of the SNN architecture, consists of three key tasks; write, read and reset. The write step involves the calculation of weighted addition of the spike events in a given time step using the crossbar, and applying the corresponding voltage to the memristor. {In order to observe whether the memristor has switched, a read phase is carried out. This can be done by a resistor divider circuit as shown in Fig. {\ref{readwritereset}}. If the memristor switched during the write phase, then the inverter output will be high. Else, it will be low.

Due to the variations in the ON,OFF resistances, the voltage at the resistor divider arrangement can vary. This may lead to erroneous identification of an occurrence of a spike, if a properly designed inverter is not present. To account for such potential errors, an inverter with a sharper characteristic curve and a controlled trip point must be used. The inverter will then identify if the resistance of the neuron memristor has gone below a certain threshold resistance (which is a spiking activity). Fig. {\ref{inv}} shows the response of the inverter we used in this work. The input voltage variations due to changes in ON and OFF states are shown in red for a standard deviation ($\sigma$) of $20\%$. We conducted $100,000$ Monte-Carlo simulations to check the number of false identifications of a spike for different values of $\sigma$. For $\sigma = 100\%$ in ON resistance, we witnessed a false identification of a spike $0.007\%$ of the time for the inverter characteristics in Fig. {\ref{inv}}.

The spiking events identified by the inverter, can be stored in buffers until the next time step. After the read phase, all the memristors will be reset to be used in the next time step. Resetting `all' the memristors is necessary since a write pulse may contribute to the growth of a conductive filament in a memristor even though it was unable to fully switch the device. Fig. {\ref{readwritereset}} (b) shows the aforementioned write, read and reset temporal activities.}

\subsection{{Implementation of the network layers using memristive crossbars}}

\begin{figure}[b!]
\centering
\vspace{-10pt}
\setlength{\abovecaptionskip}{-2pt}
\includegraphics[width=3.5in]{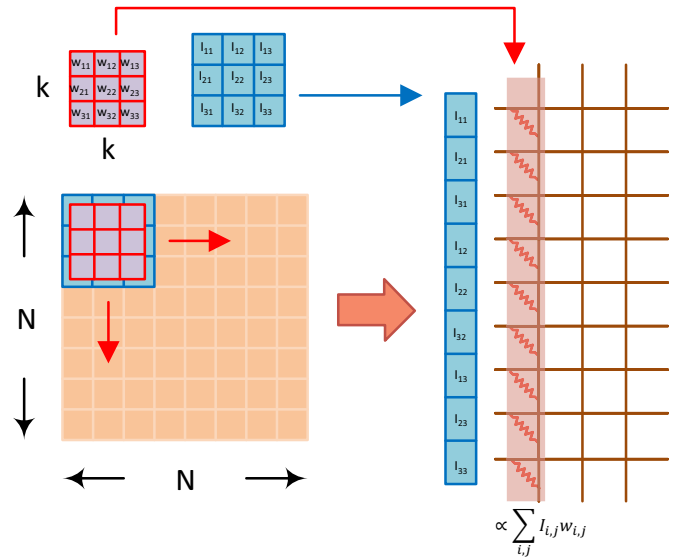}
\caption{{The dot product operation in a convolution layer. Image size is $N\times N$ and the kernel size is $k\times k$}}
\label{oneconv}
\end{figure}

{In this section, we will discuss how the fully connected, convolutional and subsampling layers are implemented using memristive crossbars. The governing operation in all these layers is the dot product, and a crossbar can be utilized efficiently for this. Fig. {\ref{xbar}} clearly shows how a fully connected layer with N input neurons and M output neurons can be implemented. The convolutional operation can be implemented as shown in Fig. {\ref{oneconv}}. As the figure illustrates, the convolution operation consists of a kernel moving over the image, getting the weighted summation in each location. A single such weighted summation can be implemented by first converting the corresponding section of the image and the kernel in to vectors and then mapping the weight values to memristors in a column in the crossbar as shown in Fig. {\ref{oneconv}}. When mapping one entire layer with $M_{in}$ input channels and $M_{out}$ output feature maps, the input must be divided into sections of $k\times k\times M_{in}$ (assuming a kernel size of $k\times k$). The crossbar should also have a similar number of conductances to obtain a single element in the output maps. Multiple such crossbars should be there to account for all aforementioned input chunks. Subsequently, all the layers in the CNN can be spatially mapped across several crossbars. Consequently, The area consumption of the entire memristive hardware required to execute the CNN can be estimated as the sum of area of all such crossbars and associated peripherals (buffers, amplifiers $etc.$). Additionally, the inference delay is estimated as the time required to sequentially propagate data across the crossbars mapped to the CNN layers. For the calculation of energy consumption, we are doing a circuit level analysis.

At the output of each crossbar, the measuring resistor, the amplifier and read, reset circuitry must be there as shown in Fig. {\ref{xbar}} and {\ref{readwritereset}}. In the architectural level, there are multiple ways of dividing the convolution operation into multiple realizable crossbar sizes {\cite{convolution},\cite{isaac},\cite{resparc}}. When dividing the operation into a number of crossbars, summing amplifiers must be incorporated to add the outputs of multiple crossbars and feed to the memristor neurons.}

{Subsampling layer (averaging) takes a similar form as above since it performs the convolution operation with a kernel size equal to the scaling factor, and all the kernel elements being equal. Therefore the subsampling layer can be implemented by using the same convolutional layer architecture. The input channels and the output maps must be selected as equal. The stride in the subsampling layer is equal to the scaling factor. In each filter set, except for the corresponding channel’s filter weights, all the other channels’ filter weights must be zero. This is due to the fact that there is only one kernel to map each input channel to a single output feature.}


\section{Results}\label{result}

In order to view the functionality of the All-Memristor based deep stochastic SNN, we have created an algorithm-device-circuit framework and tested on a standard digit recognition data set, MNIST. The {architecture selected for this work is} a convolutional neural network ($28\times28-6c5-2s-12c5-2s-10o$ \cite{deep}). The CNN structure as mentioned in section \ref{deepnn} is well known for its high recognition accuracies on complex data sets and we have chosen it for this work to show the applicability of the proposed devices on state-of-the-art neural networks. It is noteworthy that this proposed device architecture is applicable to any type of ANN (ex: fully connected) since the basic computational blocks (calculating the weighted summation) remain the same. The CNN used in this work has 2 convolutional layers followed by subsampling. Each convolutional kernel is of size $5\times5$ and there are 6 and 12 feature maps at the output of first and second convolutional layers respectively. The input image is of size $28\times28$ (an image of a digit in MNIST data set) and the pixel intensity dependent spike activity is fed to the first convolutional layer. The input spikes can also be generated by directly applying voltages to a set of memristors with an amplitude proportional to the intensity of the pixels. The memristors would then generate homogeneous Poisson spike trains proportional to the intensity of the image pixels. After each convolution layer, a subsampling with the kernel size 2x2 is present and this is simply averaging the spiking activity of few neurons. A fully connected layer appears between the second subsampled convolutional layer output and the network output. There are 10 output neurons to account for the 10 digits (classes) in the dataset. The network was trained as an ANN for $60 000$ images of handwritten digits, with the probabilistic switching curve of a memristor as the activation function of neurons, following the process mentioned in section \ref{deepnn}. The stochastic memristor neuron model is built according to the set of equations elaborated in section \ref{neuron}.

\begin{figure}[t!]
\centering
\includegraphics[width=3.5in]{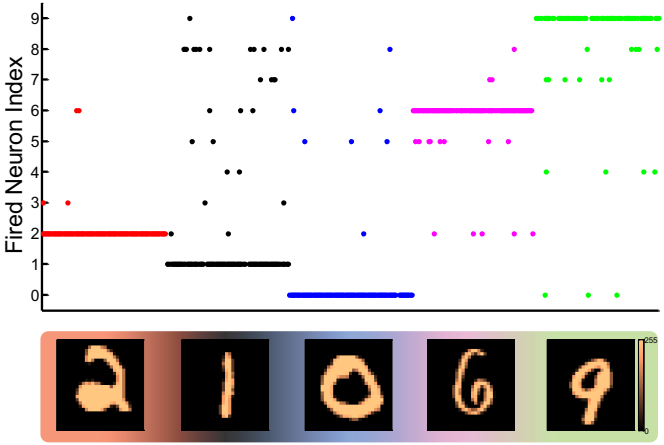}
\caption{The spiking activities of the 10 output neurons of the All-Memristor neural network over 100 time steps for randomly selected 5 images in the testing data set}
\label{spikeout}
\vspace{-10pt}
\end{figure}

The trained network was then tested on $10,000$ images of handwritten digits as a spiking neural network. Instead of evaluating the outputs of neurons as analog values, the probability of switching is determined according to the voltages applied on the memristors. These voltages depend on the weighted summation of input spikes that goes in to the neuron. We observed the spiking activities at the 10 output neurons over a 100 time steps (each including a write, read, and a reset phase) and the winner is considered as the neuron that gave the highest number of spikes during the total time interval. We obtained a classification accuracy of $97.84\%$ with a write pulse of 10ns, after detecting the spiking activity over 100 time steps. Fig. \ref{spikeout} shows the spiking activities at the 10 output neurons over 100 time steps for 5 randomly selected images from the testing data set. The accuracy we obtained for this network shows a slight degradation when considered with the baseline ANN with sigmoidal activation functions that provides an accuracy of $98.9\%$. This degradation is due to the circuit and device related considerations we took in to account while converting the ANN to an SNN. One of the reasons is the fact that we quantized the synaptic weights to suit the currently available multi-level memristors with {4-bit levels}. Another reason is the non-ideality due to the inclusion of the measuring resistor described in section \ref{deepnn}. The ANN to SNN conversion error (section \ref{neuron}) has an impact on the accuracy degradation as well.

In the next few subsections, we will discuss about the circuit level implementations and analyze the impact of different types of variations on our All-Memristor deep stochastic SNN.

\begin{figure}[b!]
\centering
\vspace{-10pt}
\setlength{\abovecaptionskip}{-2pt}
\includegraphics[width=3.5in]{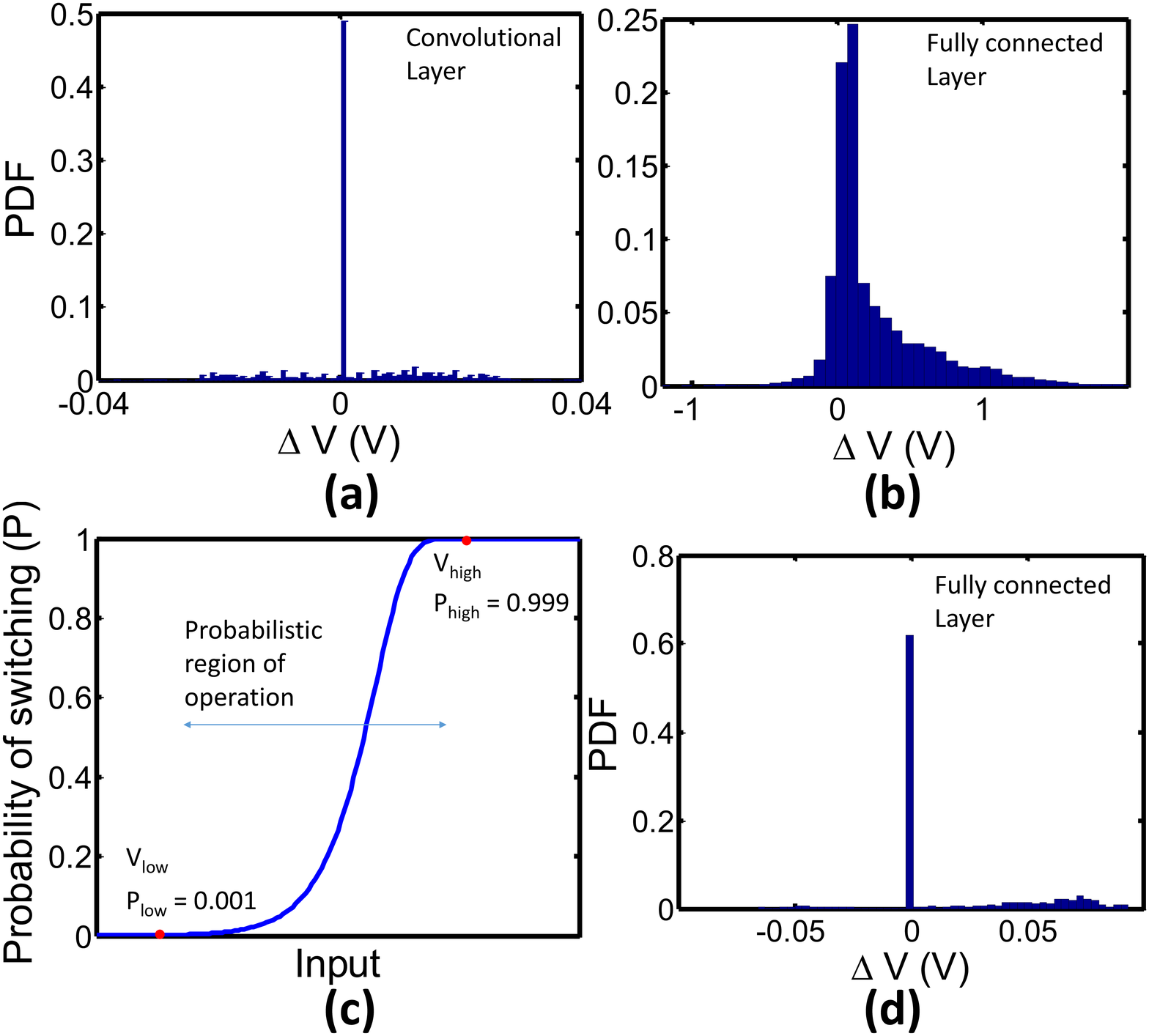}
\caption{{The variations in the input voltages to the memristors. The difference between the actual and the ideal voltage ($\Delta V$) that is being applied to a (a) convolutional layer and a (b) fully connected layer. (c) The probabilistic switching curve and the cut off voltages. (d) The $\Delta V$ when the applied voltages to the memristors are limited to $V_{high}$ and $V_{low}$}}
\label{deltav}
\end{figure}

\subsection{Circuit level simulations}

{Providing the accurate voltage to the memristive neuron is important. In order to verify that the correct voltages are being supplied to the neurons, we conducted circuit level simulations for our network. We used IBM 45nm technology node for CMOS devices. We first found the input spiking activities that must be applied to each layer, for 1000 random images in the testing data set. Then the corresponding voltages were applied to the inputs of the layers implemented in circuit level. After that, voltages applied on each memristor neuron was measured. Finally, the difference between these voltages and the actual voltages that must be ideally applied on memristors were calculated.

Fig. {\ref{deltav}} (a) and (b) show the probability density functions (PDF) of these voltage differences ($\Delta V$) for a convolutional layer and the fully connected layer of our network respectively. As the figure illustrates, for the convolutional layer, the differences in voltages fall well below $\pm40$mV ($\sim3\sigma$). However, for the fully connected layer, we noticed significant $\Delta V$ values. It is noteworthy that this does not affect the correct functionality of the network. The reason can be explained as follows.} 

\begin{figure}[t!]
\centering
\includegraphics[width=3.5in]{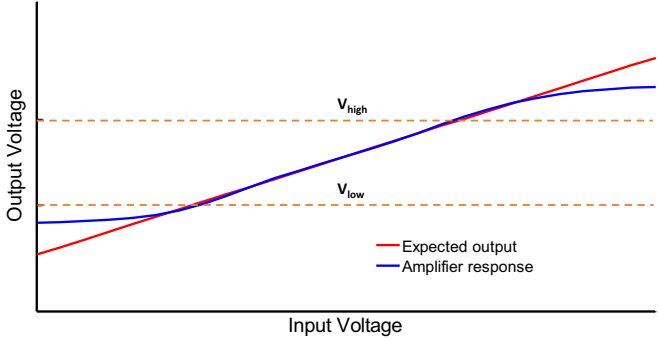}
\caption{The input output characteristics of the amplifier circuit. Note the non-linearities in the response at considerably high and low voltages}
\label{amp_resp}
\vspace{-10pt}
\end{figure}

{If a considerably high voltage is applied on the memristor, it is possible to state with higher confidence that the device will switch. For an example, if the applied voltage is larger than $V_{high}$ (Fig. {\ref{deltav}} (c)), it switches $99.9\%$ of the time. Similarly, when a sufficiently lower voltage is applied ($<V_{low}$), it can be stated with higher confidence that the device does not switch. In contrast to a convolutional layer, the output of the final fully connected layer is forced to be a set of zeros and a one during training. Due to this, the actual mapped voltages to the output neurons will be higher than $V_{high}$ or lower than $V_{low}$. Therefore, the output neuron memristors will operate on the deterministic region (Fig. {\ref{deltav}} (c)). However, the designed amplifiers may not work linearly when the inputs are very high or low. $i.e.$, as shown in Fig. {\ref{amp_resp}}, the amplifier output gets saturated. This behavior in the amplifier is appropriate since applying higher voltages to the memristor leads to higher power consumption and faster device degradation. The higher differences in Fig. {\ref{deltav}} (b) is due to such non linearities in the amplifier. We limited the applied voltage range from $V_{low}$ to $V_{high}$ to the same data we obtained for the Fig. {\ref{deltav}} (b). These cut off voltages result in $0.001$ and $0.999$ probability values respectively. This range is a good approximation given that the final classification accuracy is decided based upon a number of spikes; not just a single spike. The results in Fig. {\ref{deltav}} (d) shows the updated differences and they fall below $\pm100$mV($\sim3\sigma$). It will be shown in the next section that the network experiences only about $\sim3\%$ accuracy degradation for variations in the memristor input voltage with $\sigma =200$mV.}

\subsection {Impact of variations in the input voltages to the neurons}

\begin{figure}[b!]
\vspace{-10pt}
\setlength{\abovecaptionskip}{-2pt}
\centering
\includegraphics[width=3.5in]{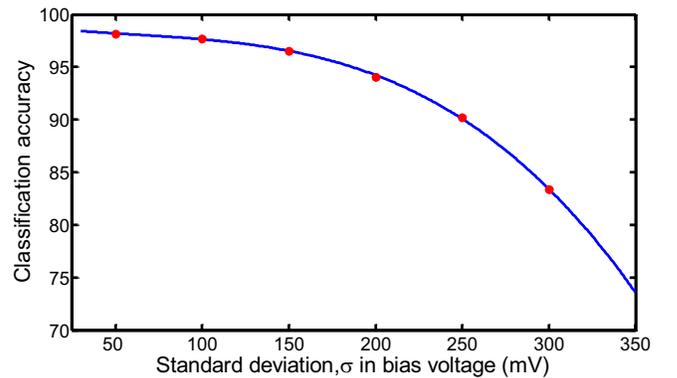}
\caption{Average classification accuracy with percentage variations in memristor neuron bias voltage. The variations follow a Gaussian distribution and independent Monte-Carlo simulations were conducted over the entire 10,000 testing image set for 50 trials.}
\label{biasvsacc}
\end{figure}

In this section we observe the effect of variations in the neuron input voltages on the classification accuracy of the network. We conduct the experiment by changing the bias voltages of the neurons. As elaborated in section \ref{deepnn} D, a bias value must be selected to account for the output probability of the neuron during a non-spiking event. $i.e.$, if no spike appears at the input of the neuron, the bias voltage is the write pulse magnitude that will be applied to the memristor neuron. We perturbed all the neuron bias voltages following a Gaussian distribution with variable standard deviations from $50$mV to $300$mV. $50$ independent Monte-Carlo simulations were conducted on all the 10000 test images. As Fig. \ref{biasvsacc} illustrates, the classification accuracy degrades by $\sim 14\%$ when the $\sigma$ is increased from $50$mV to $300$mV. The impact on accuracy increases exponentially with the increased amount of variations in the bias voltage. For an example, a 0.2V will result in just $3\%$ degradation in accuracy which is almost three times smaller compared to the $14\%$ degradation for a $300$mV variation. We would thus declare that the network is robust to variations in bias voltages less than $200$mV. The circuit simulations in the previous section show that the input voltage to the neurons are well below $100$mV.

\subsection{The impact of write time on accuracy}

As explained in section \ref{neuron}, in order to operate in the stochastic regime of a memristor, smaller write pulse widths require larger voltages and vice versa. It is however noteworthy that the switching probability curves for different pulse widths have almost the same sharpness (Fig. \ref{PvsV} (a)). The sharpness of the probability curve directly impacts the accuracy. Sharper curves will result in more classification errors. For an example, if the network was trained with a sharper curve, a slight change in a synaptic weight (due to weight quantization according to the multi-level memristors) value will result in a huge deviation at the output of a neuron to which the specific weight is connected. Fig. \ref{tvsacc} shows how the classification accuracy varies with the number of time steps considered for different write pulse widths. Higher number of time steps will result in higher accuracy. As the figure illustrates, confirming our prior argument about the sharpness, we do not see any significant relationship with respect to accuracy degradation under varying write pulse width. However, it must be noted that the bias voltage in the amplifier must be increased with the reducing write pulse width. Larger voltages might damage the device and also cause in larger power consumptions.

When the network is trained for a given write pulse width (i.e., for a particular probability curve), the variations in this write pulse width when the network is in actual operation, may cause classification accuracy degradations. In order to observe this, we perturbed the write pulse width by a certain percentage and checked the accuracy at the output of the network for all the 10000 images in the testing dataset. For a network that was trained for a 100ns write pulse width, we observed only a $0.64\%$ accuracy degradation for a $20\%$ perturbation (i.e. 20ns perturbation), and a $0.79\%$ degradation in accuracy for a $50\%$ perturbation. The same percentage perturbations were applied to a network which was trained assuming a $20$ns write pulse width. The degradation in accuracy we observed was $0.93\%$ and $1.03\%$ for $20\%$ and $50\%$ perturbation in write pulse width respectively. This explains that the network is very robust to the variations in the write pulse width.

\subsection{The impact of synaptic weight variations }

\begin{figure}[b]
\vspace{-10pt}
\setlength{\abovecaptionskip}{-2pt}
\centering
\includegraphics[width=3.5in]{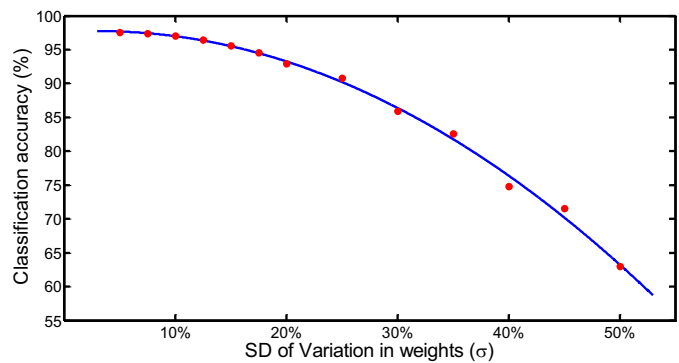}
\caption{{Average classification accuracy with percentage variations in synaptic weights. The variations follow a Gaussian distribution and independent Monte-Carlo simulations were conducted over the entire 10,000 testing image set for 50 trials.}}
\label{wvsacc}
\end{figure}

In our network, the synaptic functionality is performed by memristive crossbar arrays. Variations can be present in the memristor resistances in these crossbars due to multiple reasons including the deviations occurred during programming, the effects of temperature, and temporal drifts in resistance due to the applied small voltages. Since such process variations is a common issue {\cite{scaling}}, we tested the robustness of our memristor based SNN system to variations in synaptic weights. We perturbed all synaptic weights we obtained from our modified offline training scheme, following a Gaussian distribution with different standard deviation ($\sigma$) values. Fig. \ref{wvsacc} illustrates how the classification accuracy deviated with the increasing standard deviation (it is considered as a percentage of the weight). The accuracy degrades by $\sim4.5\%$ when the standard deviation is $20\%$. For smaller $\sigma$ values around $10\%$, the accuracy degradation is about $0.5\%$. {Despite the inherent error resiliency associated with neural networks, the accuracy degradation is significant when $\sigma = 50\%$. The work in {\cite{querlios}} also shows higher degradation in accuracy when memristors have high variations. However, the experimentally measured variability for a filament based device was as small as $\delta R/R \sim 9\%$ according to {\cite{abrupt}}. Our network is robust to variations of this magnitude.

As shown in {\cite{memvar}}, the typical write mechanisms will induce variations in multi level memristors. In order to account for this, high precision write mechanisms must be incorporated. A feedback write scheme would be appropriate to make sure that the proper value has been transferred to the memristors {\cite{write1}}, {\cite{write2}}. The work in {\cite{write1}} experimentally shows the possibility to tune the memristive device within $1\%$ accuracy degradation with respect to the desired state, within the dynamic range of the device. Furthermore, the usage of on-chip learning schemes will be helpful to account for these variations {\cite{lim}}}

Resistance variations can occur in the neuron memristor as well. However, this does not cause any significant read error at the output since the off to on resistance ratio is in the order of $10^4-10^7$ \cite{14} and the resistor divider circuit is capable of detecting this large drop with almost zero error (refer to section \ref{cctimp}). Further, as long as the amplifier output impedance is low, the write operation does not get affected by the variations in the neuron memristor.

\subsection{{Accuracy degradation of the network due to the measuring resistor $R_{meas}$}}

\begin{figure}[t]
\centering
\setlength{\abovecaptionskip}{-2pt}
\includegraphics[width=3.5in]{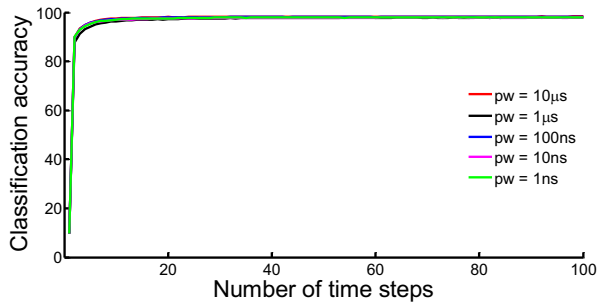}
\caption{The accuracy variation with increasing number of steps for different write pulse widths for the All-Memristor neural network}
\label{tvsacc}
\vspace{-10pt}
\end{figure}

{In a memristive crossbar, the weighted summation of a set of input voltages are given in terms of a current. This current is given by the equation ({\ref{I_vs_V}}). It must then be converted in to a voltage to feed the neuron memristor. In order to do so, a measuring resistor was incorporated as shown in Fig. {\ref{xbar}}. Due to this measuring resistor, the resultant current witnesses some non-linearities. The actual current flowing through the $R_{meas}$ can be given by the following equation}

\begin{equation}
{I_j(t)=\dfrac{V(t)\cdot[G_{1,j},G_{2,j}...G_{N,j}]^T}{1+R_{meas}\sum_{i=1}^{N}{G_{i,j}}}}
\label{I_actual}
\end{equation}

\begin{figure}[b!]
\vspace{-10pt}
\setlength{\abovecaptionskip}{-2pt}
\centering
\includegraphics[width=3.5in]{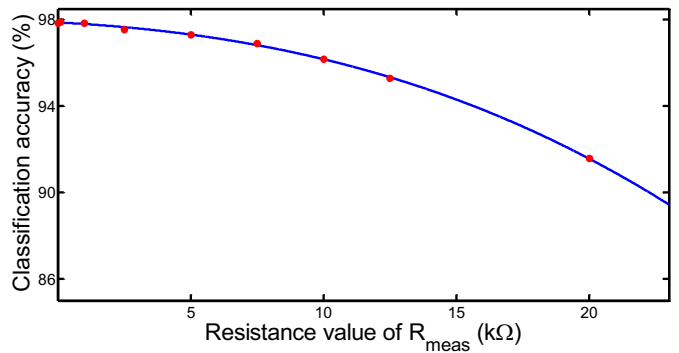}
\caption{Average classification accuracy with different values of measuring resistors ($R_{meas}$).}
\label{Rmeas}
\end{figure}

{As explained in section IV, having a smaller $R_{meas}$ with respect to the $\sum_{i=1}^{N}{G_{i,j}}$ will approximately make the current close to the inner product between $V(t)$ and $G$. In order to view the effect of the magnitude of this $R_{meas}$ towards the accuracy of the network, we evaluated the network with different $R_{meas}$ values. Fig. {\ref{Rmeas}} shows how the accuracy changes with the value of $R_{meas}$. Higher resistances result in higher accuracy degradations.}

\subsection {{The Impact of variations in neuron memristors}}

\begin{figure}[b!]
\vspace{-12pt}
\setlength{\abovecaptionskip}{-2pt}
\centering
\includegraphics[width=3.5in]{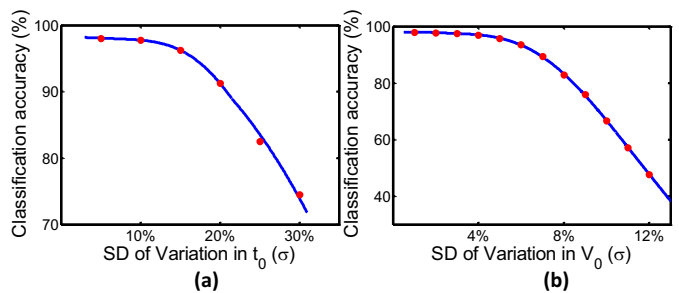}
\caption{{Average classification accuracy with percentage variations in fitting parameter (a) $\tau_0$ and (b) $V_0$ of the neuron memristors. The variations follow a Gaussian distribution with mean selected according to experimental values. Monte-Carlo simulations were conducted over the entire 10,000 testing images.}}
\label{tau0_vs_acc}
\end{figure}

\begin{figure}[b!]
\vspace{-5pt}
\setlength{\abovecaptionskip}{-2pt}
\centering
\includegraphics[width=3.5in]{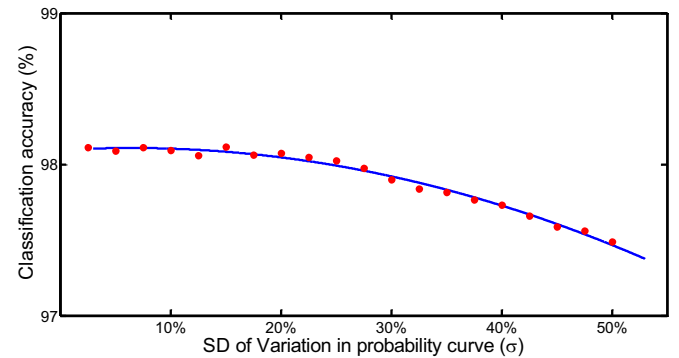}
\caption{{Average classification accuracy with percentage variations in probabilistic switching curve. Each point in the probability curve was perturbed by a $\Delta P$ amount following a Gaussian distribution. The accuracy of the network was measured over 5000 images on the testing data set}}
\label{P_vs_acc}
\end{figure}

{The switching probability of a neuron memristor depends on the two fitting parameters ${\tau_0}$ and ${V_0}$ (equation ({\ref{tau0}})). Due to variations, different memristors in the network can have different ${\tau_0}$ and ${V_0}$ values. In order to view the effect of the variations in these parameters towards the accuracy of the network, Monte- Carlo simulations were conducted. The parameters were perturbed following a normal distribution while considering the experimental values {\cite{14}} as the mean. As Fig. {\ref{tau0_vs_acc}} (a) illustrates, the effect of the variations in ${\tau_0}$ (up to $20\%$) towards the accuracy of the network is small ($\sim5\%$). In contrast, the accuracy of the network is sensitive to the variations in ${V_0}$. When the percentage variation of ${V_0}$ goes above $6\%$, the accuracy degrades significantly (Fig. {\ref{tau0_vs_acc}} (b)). When a particular memristor has a higher ${V_0}$ value, the neuron corresponding to that memristor has a lower probability of switching than what it was designed for, and vice versa. However, the experimental studies have shown that the thickness can affect the ${V_0}$ value and this effect is not significant. For example, it has been shown that when the thickness of the memristor was scaled by a factor of 2 (increased by $100\%$), the ${V_0}$ increases only by $\sim35\%$ {\cite{14},\cite{20}}. Therefore, it can be argued that a $5-10\%$ variation in dimensions will not cause significant classification accuracy degradation.

Cycle-to-cycle variations in ON and OFF resistances of the neuron memristors can occur as well. However, these variations have already been accounted for while experimentally obtaining the switching probability curve. Furthermore, the memristors degrade after a certain number of set-reset cycles. This can impact the probabilistic switching curve and thus the accuracy of the network. In order to view the impact of these changes, we perturbed the probabilistic switching curve by a small amount ($\Delta P$) and simulated the network on 5000 images for 100 iterations. As Fig. {\ref{P_vs_acc}} illustrates, the network is robust for variations in the probabilistic switching curve. However, sufficient data is not available in literature to exactly represent the effect of memristor degradation on the probabilistic switching curve (for the particular device we selected).}

\subsection{Delay and energy consumption of the SNN}

\begin{figure}[b]
\vspace{-12pt}
\setlength{\abovecaptionskip}{-2pt}
\centering
\includegraphics[width=3.5in]{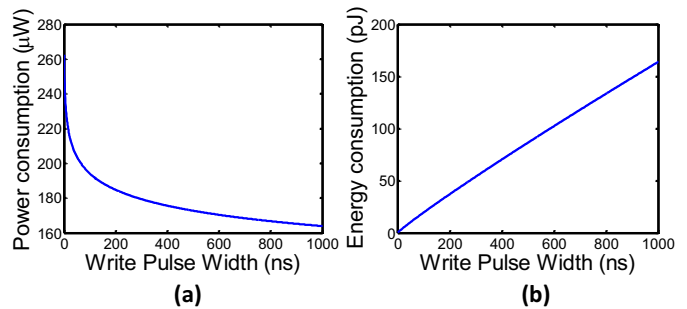}
\caption{(a) The neuron power consumption and (b) energy consumption for different write pulse widths. The experiment was for an amplifier output voltage that corresponds to a switching probability of 0.5 for the selected write pulse width. Even though the power consumption reduces with the increasing pulse width, the energy consumption grows.}
\label{powerandenergy}
\end{figure}

As we noted in Fig. \ref{PvsV}, in order to get the same switching probability for a memristor, the lower write pulse widths require higher write pulse magnitudes. Since the relationship between the energy and the write pulse width is not quite intuitive, we calculated the energy consumption of a single neuron for different write pulse widths. The results are summarized in Fig. \ref{powerandenergy}. Here we assumed a spiking activity of $0.5$ at the input. The results suggest that larger pulse widths result in larger energy consumption. This is due to the exponential relationship between the voltage and pulse width. For example, if the write pulse width must be reduced from $1\mu s$ to $100ns$ to achieve faster operation, the required voltage increment is just ~500mV and the energy consumption would be better than the memristor operating in $1\mu s$ (even though the power consumption reduced). 

\begin{table}[t]
\renewcommand{\arraystretch}{1.3}
\centering
\caption{Device simulation parameters}
\label{table1}
\begin{tabular}{c|c }

\hline \hline

Parameters & Values \\ \hline
On resistance ($R_{on}$) & $500$k$\Omega$ \cite{spec} \\
On/Off ratio & $10^3$ \cite{spec} \\ 
Thickness of the insulation, $t_{a-Si}$ & $60nm$ \cite{14} \\
Fitting parameter $V_0$ & $0.22$ \cite{14,20} \\
Fitting parameter $\tau_0$ & $2.85\times 10^5$ \cite{14,20}\\
Crossbar operating voltage & $1V$ \cite{spec}\\
\hline 

\end{tabular}
\vspace{-5pt}
\end{table}
When considering the energy consumption of the entire system per image classification, the number of time steps (write, read, and reset cycles) plays an important role. The accuracy of the network increases with the number of time steps over which the winning neuron is decided (Fig. \ref{tvsacc}). A reasonable accuracy (above $96\%$) can be reached within 10 time steps as shown in Fig. \ref{tvsacc}. However, for more complex datasets, the convergence time can be much higher ($\sim50$ time steps) {\cite{binghan}} due to the fact that the data sets are much bigger, and more number of neurons are required to increase the accuracy \cite{rbfn}. Hence, for our energy comparison with CMOS baseline, we conservatively choose 50 time steps for SNN inference.

In order to calculate the energy consumption of the whole network, we used SPICE simulations in IBM 45nm technology. We considered the average spiking activities for the images in the testing data set. The crossbar voltages must be selected appropriately (depending upon the type of memristors used) so that the drift in the resistance values over time is minimal. All the important parameters involved in this work are included in Table 1. The energy required for a single write to the neuron memristor (along with the amplifier) for a switching probability of $50\%$ is $249$fJ. A single reset and a single read operation consumes $\sim 500$fJ and $\sim 1.4$fJ amount of energy respectively. {The average energy per image classification was $115$nJ. Note that in our energy estimation for CNN execution (both memristive and CMOS hardware) we assume that input data is available in the form of spikes (two voltage levels). Such Poisson rate-encoded spike streams can be potentially obtained from event-driven sensors {\cite{sensor}}. The energy consumption due to the event-sensor operation would be a common component to both the memristor and CMOS implementations. The aforementioned energy number includes the energy associated with the peripheral buffer read and write, crossbars, and the read-write-reset of the neurons.}

We observed that the energy of the crossbar is the dominant component and this is justifiable due to the fact that the number of synapses are orders of magnitude larger than the number of neurons. For example, the last fully connected layer of the network has 1920 synapses and the number of neurons are only 10. This is a $\sim \times 200$ difference and thus we state the results are justifiable. The second dominant energy component is from the reset operation. This is because of the fact that the reset must be conducted in the deterministic region of operation of a memristor. That is, a high enough voltage pulse must be used to ensure that the device has turned off. Since the resistor value is now lower, the energy consumption is larger for this step. To address this, feedback reset mechanisms can be incorporated \cite{17}. This will allow the operation in lower voltage stochastic regime with some feedback control circuitry that conditionally gets activated. Furthermore, a novel stochastic volatile memristor has been proposed in \cite{26} as a true random number generator. It has been shown experimentally that once this memristor is turned on, it returns to its off state after a small duration of time ($\sim 100\mu s$) eliminating the requirement to force reset as we propose in this work. The write voltage is also lower (0.5V, for a $300\mu s$ pulse) in this device when compared with $Ag/a-Si/p-Si$ (3.3V for $1ms$ pulse \cite{14}) and $TiO_2$ devices. This may even eliminate the requirement of high gain amplifiers that consume power. We thus argue that there are other types of memristor devices that can allow energy efficient implementation using the architecture we propose here. Our goal is addressing the applicability of electric field driven memristors in general for {deep stochastic SNN}. Therefore we conducted the energy calculations without the lack of generality.

\begin{table}[t]
\renewcommand{\arraystretch}{1.3}
\centering
\caption{{Area and delay of CMOS and memristor based implementations}}
\label{table2}
\begin{tabular}{c|c c|c|c }

\hline \hline

Implementation & \multicolumn{2}{c}{Area ($\times 10^{-3}\mu$m$^2$)} \vline & Delay & \makecell{Area$\times$Delay \\ (ns mm$^2$)}\\ \hline
\multirow{ 3}{*}{Memristor based} & Crossbars & $2895$ & \multirow{ 3}{*}{$210$ns} & \multirow{ 3}{*}{$652$} \\
& Neurons & 154 & & \\
& Buffers & 56 & & \\ \hline
CMOS based & \multicolumn{2}{c}{$190$} \vline & {$28\mu$s} & $5320$ \\ 
\hline 

\end{tabular}
\vspace{-5pt}
\end{table}

\subsection{Comparison with CMOS implementation}

For the purpose of comparison of our work, we used the CMOS spiking network baseline proposed in \cite{resparc}. The weight data are stored in SRAM. Subsequently, neurons in the CNN are temporally scheduled on the computation core comprised of FIFOs and neuron units. Each neuron computation involved moving data (input and weight) from SRAM to FIFOs and moving the computed outputs from neuron units back to SRAM when computation is completed. The energy of the CMOS design along with the data fetching energy from the memory, is $\sim 736nJ$ (with $130$nJ for memory accessing, $64$nJ for buffers, and $542$nJ for neurons). The energy number is for iso-number of time steps as our proposed All-Memristor network (50 steps). This is approximately 6.4 times larger than the energy consumption of the proposal. However, we would like to point out that memristor neurons degrade faster over time when compared with CMOS neurons, even though the endurance of memristors is significant (up to $10^{12}$ set-reset cycles). Larger operating voltages may speed up this degradation process as well \cite{28}. The on-off resistance ratio of a memristor changes after a certain number of write cycles and may have different switching probability curves other than the one used for training the network. This will lead to lowered classification accuracies as explained in section \ref{result} F. However, retraining might help in regaining some lost accuracy but the feasibility of this is debatable. 

In order to find the delay and area of our design, we divided the computation in to $128\times128$ memristive crossbar arrays following the procedure mentioned in section \ref{cctimp} A. The total area of the design (including crossbars, neuron memristors, buffers, amplifiers, and inverters) was $\sim3mm^2$ (Table {\ref{table2}}). The crossbar cell size was assumed to be $100F^2$ {\cite{delayne}}. In our design, for a write time of $10$ns {\cite{14}}, a single step takes $42$ns (crossbar access time of $ \sim10$ns {\cite{delay1}}{\cite{delayne}}, read and buffer time $ \sim2$ns, reset time $ \sim20$ns). The latency for a single spike is $210$ns (the propagation time through the full network explained at the beginning of section \ref{result}). The latency for a single spike for the spiking CMOS architecture {\cite{resparc}} takes $28 \mu s$ for the same network. In order to fairly compare the two implementations, we are estimating the area $\times$ delay product. The product for our proposal is $652$nsmm$^2$. For the CMOS implementation, the value is $5320$nsmm$^2$ (which is $\sim 8\times$ bigger and thus worse).

\section{Conclusion}
Memristive switching devices have shown to be promising candidates for an enormous array of applications including logic, memory and neuromorphic computing. However, their inherent stochasticity has given life to reliability concerns. Numerous mechanisms involving larger write pulse widths, larger operating voltages, or feedback architectures have been proposed to drive these highly stochastic devices to their deterministic operating regime. As a result, we have to pay in terms of larger power consumption. This work is an effort of exploring an avenue where such stochasticity can be embraced rather than eliminating, with the goal of reducing power consumption. The proposal is embedding the functionality of a stochastic neuron to a memristor while representing the synaptic weights by a memristive crossbar to build the``All-Memristor Deep Stochastic SNN''. 

We tested the functionality of the network using the MNIST handwritten digit data set and witnessed a very low accuracy degradation ($\sim 1\%$) when compared with the deep {ANN} baseline. The design space of the network was estimated by applying variations and we observed that our proposal is robust to variations in the synaptic weights ($\sigma < 20\%$), neuron bias voltages ($<200$mV), probabilistic curve, and the write time durations ($\sim 50\%$ of the pulse widths). The steepness of the activation function of a neural network affects the accuracy of the output and makes it less robust to variations. The constant steepness of the switching probability curve of a memristor (the probabilistic activation function) over different write times gives more flexibility for the memristor to be utilized in platforms with different speed limits without creating any accuracy degradation. 

Smaller write pulse widths require larger voltages to bring the memristor to its stochastic region of operation. However, the required increment in voltage magnitude to operate $10$ times faster is small ($<500$mV) leading to lower energy consumption at the neuron in fast operating platforms. Furthermore, the total energy consumption of the proposal is $6.4$ times smaller, and the area$\times$delay product is $8$ times smaller when compared with the digital CMOS counterpart.


%



\section*{Acknowledgment}

The authors would like to thank C. M. Liyanagedera of Purdue University for his help in discussing and implementing the analog circuitry and for reading the manuscript. The work was supported in part by Center for Brain Inspired Computing (C-BRIC), a MARCO and DARPA sponsored JUMP center, by the Semiconductor Research Corporation, the National Science Foundation, Intel Corporation and by the Vannevar Bush Fellowship.

\ifCLASSOPTIONcaptionsoff
\newpage
\fi



\bibliographystyle{IEEEtran}
%

%








\end{document}